\documentclass[]{spie}  %>>> use for US letter paper
\usepackage{amsmath,amsfonts,amssymb}
\usepackage{graphicx}
\usepackage{tabularx}
\usepackage[colorlinks=true, allcolors=blue]{hyperref}
\hypersetup{final}
\usepackage{wrapfig}
\usepackage{url}
\usepackage{placeins}

\title{How well do U-Net-based segmentation trained on adult cardiac magnetic resonance imaging data generalise to rare congenital heart diseases for surgical planning?}

\author[a]{Sven~Koehler}
\author[b]{Animesh~Tandon}
\author[b]{Tarique~Hussain}
\author[d]{Heiner~Latus}
\author[c]{Thomas~Pickardt}
\author[c,e]{Samir~Sarikouch}
\author[f]{Philipp~Beerbaum}
\author[b]{Gerald~Greil}
\author[g]{Sandy~Engelhardt}
\author[a]{Ivo~Wolf}

\affil[a]{Department of Computer Science, Mannheim University of Applied Science, Germany}
\affil[b]{Department of Pediatrics (Cardiology); Department of Radiology, UT Southwestern Medical Center, Dallas, TX, USA}
\affil[c]{German Competence Network for Congenital Heart Defects; DZHK (German Centre for Cardiovascular Research), Berlin, Germany}
\affil[d]{Department of Paediatric Cardiology and Congenital Heart Defects, German~Heart~Centre~Munich, Germany}
\affil[e]{Department of Cardiothoracic, Transplantation and Vascular Surgery, Hannover~Medical~School, Hannover, Germany}
\affil[f]{Department of Pediatric Cardiology and Pediatric Intensive Care, Hannover~Medical School, Hannover, Germany}
\affil[g]{Working group Artifical Intelligence in Cardiovascular Medicine, Heidelberg University Hospital, Heidelberg, Germany}

\begin{document} 
\maketitle
\begin{abstract}
%Purpose:
Planning the optimal time of intervention for pulmonary valve replacement surgery in patients with the congenital heart disease Tetralogy of Fallot (TOF) is mainly based on ventricular volume and function according to current guidelines. Both of these two biomarkers are most reliably assessed by segmentation of 3D cardiac magnetic resonance (CMR) images. In several grand challenges in the last years, U-Net architectures have shown impressive results on the provided data. However, in clinical practice, data sets are more diverse considering individual pathologies and image properties derived from different scanner properties.  Additionally, specific training data for complex rare diseases like TOF is scarce.

For this work, 1) we assessed the accuracy gap when using a publicly available labelled data set (the Automatic Cardiac Diagnosis Challenge (ACDC) data set) for training and subsequent applying it to CMR data of TOF patients and vice versa and 2) whether we can achieve similar results when applying the model to a more heterogeneous data base.

Multiple deep learning models were trained with four-fold cross validation. Afterwards they were evaluated on the respective unseen CMR images from the other collection. Our results confirm that current deep learning models can achieve excellent results (left ventricle dice of 0.951$\pm{0.003}$/0.941$\pm{0.007}$ train/validation) within a single data collection. But once they are applied to other pathologies, it becomes apparent how much they overfit to the training pathologies (dice score drops between 0.072$\pm{0.001}$ for the left and 0.165$\pm{0.001}$ for the right ventricle). 
\end{abstract}

\section{\hspace{14pt}Motivation}
\label{sec:Motivation}
Tetralogy of Fallot (TOF) is the most common cyanotic congenital heart disease, affecting approx. 1:2500 babies (cf. European TOF guidelines\cite{tof_guidelines_2010} and American cardiology congress guidelines\cite{accaha_2008}). Currently, the initial repair in infancy is very successful, but often leaves the patient with progressive heart valve insufficiency. Over time, this can lead to dysfunction of the left and right heart chambers (ventricles) in about 25\% of adults, ultimately causing heart failure. Current medical guidelines suggest replacing the affected heart valve by a prosthetic valve to counteract ventricular dysfunction. However, it remains extremely difficult to decide on the optimal point in time for replacement, which is often done in patient’s adolescence, since the durability of artificial prostheses is limited. Therefore, TOF patients have to undergo several surgical replacements in their lifetime, and each surgery is associated with high risks.

Current guidelines for the timing of a heart valve replacement, more specifically of pulmonary valve replacement, are mainly based on enlarged ventricular volumes and depressed ventricular function (cf. \cite{tof_guidelines_2010}, \cite{accaha_2008}). Technically, these information can be obtained by segmentation of cardiac magnetic resonance (CMR) images. 

Current deep learning models have been shown to be capable of fully-automatic segmentation of CMR data sets \cite{Bernard2018a}. For example, in a huge international challenge for CMR, held at the Medical Image Computing and Computer Assisted Intervention (MICCAI) conference in 2017, called ACDC, deep learning methods outperformed previous methods. The best approaches use deep convolutional architectures. Most of them build on the U-Net architecture (introduced by Ronneberger et al\cite{Ronneberger2015}).

Training one deep learning model per possible pathology requires manual annotation of hundreds of images per pathology which is a time consuming task. Especially the field of congenital heart diseases suffer from a lack of clinical cases which stands in contrasts with the huge clinical heterogeneity of the given data. In this paper, we study how well models trained on the publicly available ACDC data generalise to the case of TOF patients.

\section{\hspace{14pt}Material and Methods}

\subsection{Datasets}
\label{datasets}
To address the clinical and technically motivated questions stated in section \ref{sec:Motivation}, two multi-centric datasets were analysed. Table \ref{table:GCN_overview} provides an overview of the two used datasets.

\begin{table}[hbt]
\centering
\caption{Overview of the GCN and ACDC datasets.}
\resizebox{.9\textwidth}{!}{%
    \begin{tabular}{c|c|c|c|c|c}

        Name of Cohort & Years collected & \# Patients & Heart centers & Pathologies\\ 
        \hline
        GCN dataset & 2005 - 2008 & 203 labeled of total 406 & 14 German heart centers & 4 TOF sub-pathologies\\ 

        ACDC dataset & 2017 & 100 labeled of total 150 & 1 center in Dijon & 4 pathologies + 1 healthy group\\

    \end{tabular}%
}
\label{table:GCN_overview}
\end{table}

The ACDC dataset covers adults with normal cardiac anatomy and function and the following four cardiac pathologies: systolic heart failure with infarction, dilated cardiomyopathy, hypertrophic cardiomyopathy and abnormal right ventricular volume. Each pathology is represented by 20 labeled patients and has labels for the \textit{right ventricle} (RV), \textit{left ventricle} (LV) and the \textit{left ventricular myocardium} (MYO). Each label is given at the \textit{end systolic} (ES) and \textit{end diastolic} (ED) time step.

\begin{figure}[ht]
\centering
  \includegraphics[width=.9\textwidth]{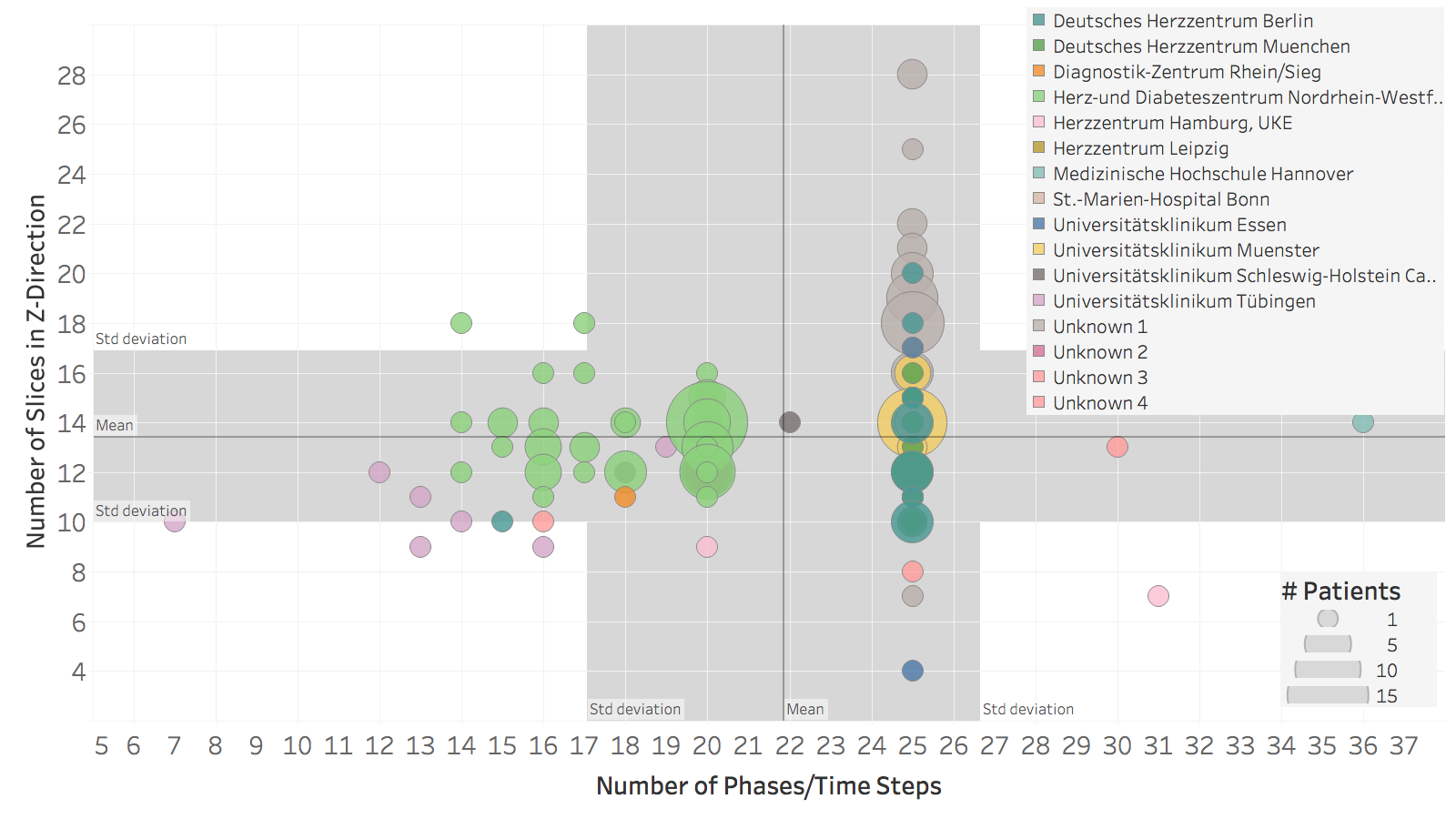}
  \caption{\label{fig:volume_distribution}Distribution of the number of phases and axial slices of the CMR images, grouped by acquisition site. The x-axis represents the number of available timesteps per patient and the y-axis the number of slices in the axial patient direction. All patients with the same in-plane resolution and the same clinical origin are grouped into one circle, the circle size shows the number of patients. The circle colour groups the clinical origin of these patients.}
\end{figure}

The German Competence Network for Congenital Heart Defects (GCN) data set is derived in the follow up study of post-repair Tetralogy of Fallot\cite{follow_up_study_2019} from multi-center collaborations of 14 German heart centers (cf. figure \ref{fig:volume_distribution}) and contains clinical, ECG and CMR data from patients with repaired TOF. This work relies only on the post surgery CMR images of the GCN study.
The ACDC dataset represents adult hearts, while the GCN subset considered in this study consists of CMR images of adolescents with the mean age of 17$\pm{8}$ years.
The RV, LV and MYO of 203 patients in the GCN dataset were labelled manually by clinical experts. Each patient has five labeled time steps, reflecting the five heart phases: end diastolic, mid systolic, end systolic, peak flow and mid diastolic phases, which makes this data set quite unique compared to other segmentation datasets. The initial analysis of the data set suggests that the provided CMR data is very heterogeneous considering size, resolution, scanner-model and number of available cardiac phases (cf. figure \ref{fig:volume_distribution} and \ref{fig:volumes_per_scanner}).

\subsection{Architecture}
\label{sec:architecture}
The segmentation network has a U-Net architecture with four down-sampling and four respective up-sampling blocks (based on the original U-Net from Ronneberger et al \cite{Ronneberger2015}) with some small adjustments: batchnormalization layer after each non-linear activation, higher dropout rates, ELU activation instead of RELU, Dice loss function, data augmentation and no resampling to force the network to learn different physical representations to increase the generalisation of the network. 

\begin{figure}[htb]
  \centering
  \begin{minipage}[b]{0.45\textwidth}
  \label{fig:network_blocks}
    \includegraphics[width=\textwidth]{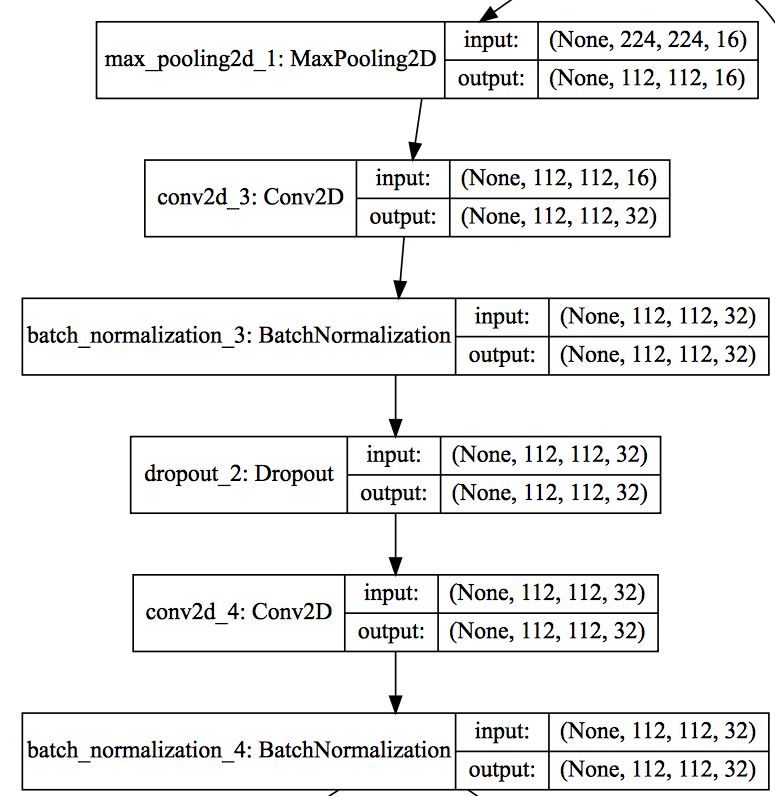}
    \caption{\label{fig:downsampling_block}Example representation (first block) for one of the four down-sampling blocks used in this U-Net.}
  \end{minipage}
  \hfill
  \begin{minipage}[b]{0.45\textwidth}
    \includegraphics[width=\textwidth]{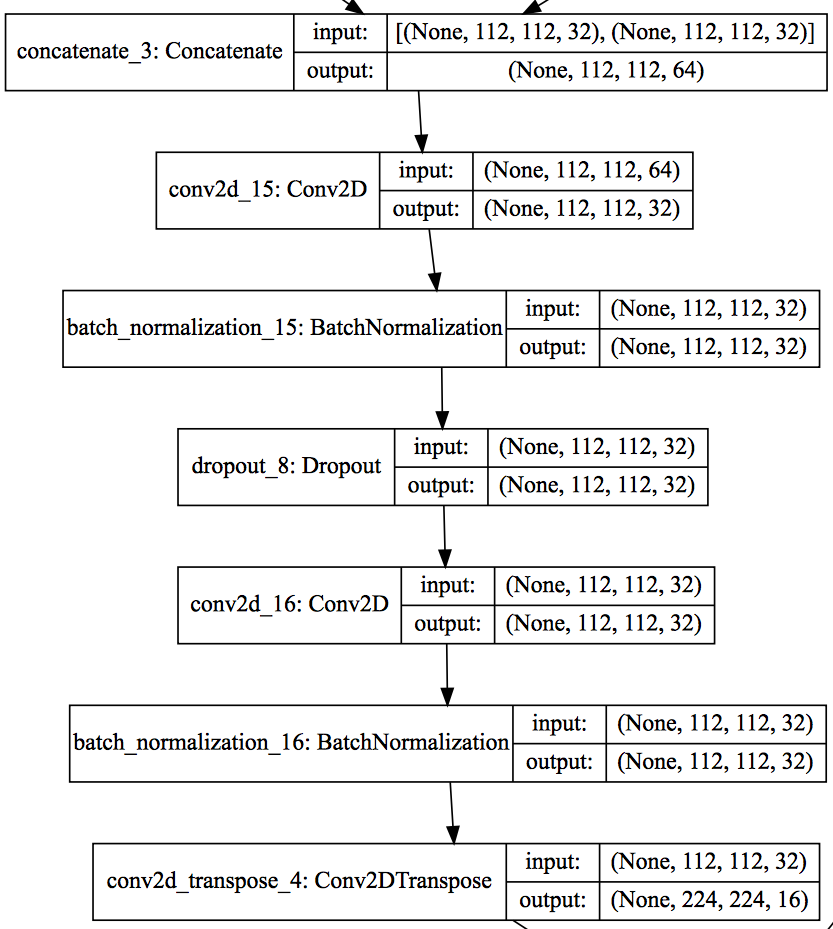}
    \caption{\label{fig:upsampling_block}Example representation (last block) for one of the four up-sampling blocks used in this U-Net.}
  \end{minipage}
\end{figure}

Figure \ref{fig:downsampling_block} and \ref{fig:upsampling_block} exemplary show one down- and up-sampling block. The ELU activation which is applied to each convolutional layer is not listed to keep it readable.
The encoding part has four down-sampling blocks, each of them consists of a 2x2 max pooling layer, followed by a 3x3 convolution layer, a batchnormalization layer, a dropout layer and again a  3x3 convolution layer followed by a batchnormalization layer. All convolution layers use zero padding. The dropout rate at the shallow encoding and decoding layers were set to 0.3, this rate increased up to 0.5 in the deeper layers. The decoding part consists of the same amount of four up-sampling blocks, each of them has a concatenate layer, a 3x3 convolution layer, a batchnormalization layer,  a dropout layer, 3x3 convolution layers, a batchnormalization layer and a transpose layer with stride = (2,2) and zero padding. Contrary to the original recommendations from Ioffe and Szegedy\cite{ioffe_batch_nodate} it turned out that the batchnormalization layer works better if it is attached after the non-linear ELU activation function. To work with categorical labels a sigmoid activation was applied on the last layer of the U-Net. Several activation functions, optimizers and parameters were tested for this network, but they did not produce reasonable different results. For the final evaluation we used the ADAM optimizer with a factorised decreasing learning rate. The trainings setup and parameters are described in section \ref{sec:training}. The network input was set to (32, 224, 224, 1). Within this study a 3D version of this U-Net was also implemented, but the 2D version beat the 3D network in all tests. Because of that, the following sections will only describe the results for the 2D model. The U-Net implementation from Isensee et al.\cite{Isensee2018} was also applied to the GCN dataset. The model architecture of the original paper needed to be modified to work with the non-isotropic spatial resolution of the CMR images. Unfortunately the segmentation performance of this modified Isensee-model was outperformed by the own 2D model, which could be related to the modified layers and the case that the 3D volumes of GCN dataset are not well aligned in the axial direction.

\FloatBarrier

\subsection{Losses and metrics}
\label{sec:metrics_and_losses}

The following loss functions and their combinations have been tested: 
\begin{itemize}
    \item Binary cross-entropy (BCE)\\
    \begin{equation}
        BCE = -\frac{1}{N} \sum_{i=1}^{N}g_{i} \cdot log (p_{i}) + (1 - g_{i})\cdot log(1 - p_{i}) \label{eq:bce}
    \end{equation}\\
    where the sums are calculated over the N voxels of the predicted mask volume (softmax output of the network) $p_i$ $\in$ P and the ground truth binary mask volume $g_i$ $\in$ G.\\

    \item Weighted cross-entropy (WCE)\\
    \begin{equation}
        WCE = -\frac{1}{N} \sum_{i=1}^{N} w_{k}\cdot(g_{i} \cdot log (p_{i}) + (1 - g_{i})\cdot log(1 - p_{i})) \label{eq:wce}
    \end{equation}\\
    where $w_{k}$ is a vector with one multiplicative coefficient per class $k$. Values smaller than 1 reduce, while values greater 1 increase the weighting of this class. This loss was introduced by Dalca et al.\cite{dalca_anatomical_2018}.\\
        
    \item Jaccard distance loss (JDL)\\
    \begin{equation}
        JDL = 1 - \frac{\sum_{i=1}^{N} g_{i} \cdot p_{i}  + smooth}{\sum_{i=1}^{N} g_{i} + \sum_{i=1}^{N} p_{i} - \sum_{i=1}^{N} g_{i} \cdot p_{i} + smooth}\label{eq:jac_distance}
    \end{equation}\\
    with $smooth$ set to 1 to avoid dividing by zero.\\
    
    \item Soft Dice loss (SDL)\\
    \begin{equation}
        SDL = 1 - \frac{1}{C} \sum_{i=1}^{C} DSC_{i}\label{eq:dsc_loss}
    \end{equation}
    where C represents the foreground classes and $DSC_{i}$ the binary soft Dice coefficient (eq. \ref{eq:dice_class}) for one class. The original SDL by Milletari et al.\cite{milletari2016v} used the squared sum, which did not work well in our experiments. Because of this, the plain sum, as described by Drozdzal et al. \cite{drozdzal_importance_2016}, is used.
\end{itemize}

The WCE learned the fastest. Overall the SDL produced the best results. All loss functions worked better by ignoring the background channel during training. A learning rate scheduler with a patience of five epochs, an initial learning rate of $\mathrm{10}^{-3}$, a decreasing factor of 0.5 and a minimal learning rate of $\mathrm{10}^{-8}$ was used to adapt the learning rate, trainings with no change in the loss for more than 10 epochs were stopped. 

To measure the performance of the models across the two datasets the Dice coefficient was used. Equation \ref{eq:dice_class} gives the binary soft Dice coefficient which is used to define the model performance per label. 
\begin{align}
\begin{split}
  DSC_{class}     & = \frac{2 \cdot \sum_{i=1}^{N} g_{i} \cdot p_{i}  + smooth}{\sum_{i=1}^{N} g_{i} + \sum_{i=1}^{N} p_{i} + smooth}\label{eq:dice_class}
\end{split}
\end{align}
In the following this will be referred to $DSC_{class}$. To interpret the performance across all labels within one image the class-wise averaged dice coefficient is calculated for the foreground classes. The overrepresented background class is ignored to reflect the performance of the labels of interest. In the following, this metric will be referred as $DSC_{labels}$ and is used to evaluate the overall performance drop between the train/test splits and the unseen other dataset. All losses were able to produce a $DSC_{labels}$ above 85\%.

\subsection{Pre-Processing}
\label{sec:preprocessing}
The following pre-processing/augmentation steps have been applied on each batch:
\begin{itemize}
\label{list:preprocessing_steps}
    \item No resampling to get different sizes - it turned out that the network learned slower but generalised better to unseen datasets if the data is not resampled. By this, the network is forced to learn the segmentation features in any size.
    \item Using the grid distortion method from the library Albumentations\cite{albumentations} with a probability parameter of 80 \% during training. The number of distortion steps are kept to the standard value of 10. The default behaviour of this function is to distort with linear interpolation on the images and with nearest neighbour interpolation on the masks. This augmentation method increased the robustness of the model to unseen data a lot. Other augmentation methods (random zoom, rotation, shift and cropping) are tested but the combination of no resampling (c.f. first preprocessing point) and the grid distortion performed the best.
    \item Cropping to square - to avoid distortions in subsequent resizing steps.
    \item Center cropping - if image size is bigger than the network input shape, else resize image.
    \item Resize to proper network size if cropping is not possible - bi-linear interpolation and anti-aliasing for the images, nearest neighbour interpolation and anti-aliasing for the masks.
    \item Pixel value clipping - value clipping based on the .999 quantile to improve the pixel intensity distribution. Some images had outliers with pixel values near to 20.000 while 99.9 \% of the pixel values are below 2000.
    \item Normalise pixel values - min/max-normalise each image to scale the pixel values of different scanners into the same range. Scaling to zero mean and one standard deviation was also tested, but this led to worse results.
\end{itemize}

\subsection{Training}
\label{sec:training}
Both datasets were shuffled and split in four folds. One model was trained per fold, which results in four models per dataset. Neither the model architecture/parameters nor the preprocessing and training parameters were optimised at this step to avoid any parameter based advantages/disadvantages for each of the datasets.
The ACDC dataset was split with respect to the five pathologies. This resulted in 15 patients per pathology per training fold and five patients per pathology per test fold. All together this led to 75 patients in each training fold with 1413$\pm{8}$ slices and 25 patients in each test fold with 471$\pm{8}$ slices. Figure \ref{fig:ACDC_data_splitting} in the Appendix provides more details on the pathology based ACDC splits.
The GCN dataset was randomly split. This resulted in 152 patients per training fold with 10544$\pm{177}$ slices and 51 patients per test fold with 3514$\pm{177}$ slices.

The following two experiments were evaluated. In experiment one four models were trained on the ACDC dataset. Afterwards they were validated on the corresponding unseen test splits and the unseen pathologies from the GCN dataset. In experiment two four models were trained on the GCN trainings splits and validated on the corresponding unseen GCN test splits and the unseen ACDC dataset.
Early stopping with a patience of 10 epochs was applied to the learning process. The initial learning rate was set to 0.001, this learning rate was decreased by a factor of 0.5 (minimal learning rate = $1\times10^{-8}$) after five epochs without any gain in the loss.

\subsection{Finetuning}
After the train/test/unseen pathology gaps for both datasets were defined, three methods for closing the gap between the public available dataset and the unseen TOF pathology were investigated. Within the first method an increasing amount of GCN patients were successively added to the ACDC trainings dataset, for each extended dataset a new model was trained without changing the model-parameters. For the second and third approach, one trained ACDC model from the crossvalidation (cf. section \ref{sec:training}) was randomly chosen as a baseline model and finetuned. For the second method, the baseline model was finetuned with the ACDC trainings split plus an increasing amount of GCN patients, whereas for the third approach the baseline model was finetuned only on an increasing amount of GCN patients. For each method ten models were trained each with a different amount of additional GCN patients (5 to 150 patients). Each model was then evaluated on the ACDC trainings split, the unseen ACDC test split and the unseen rest of the GCN dataset.

\section{\hspace{14pt}Results}

\subsection{Generalisation gap}
\begin{figure}[ht]
\centering
  \includegraphics[width=.9\textwidth]{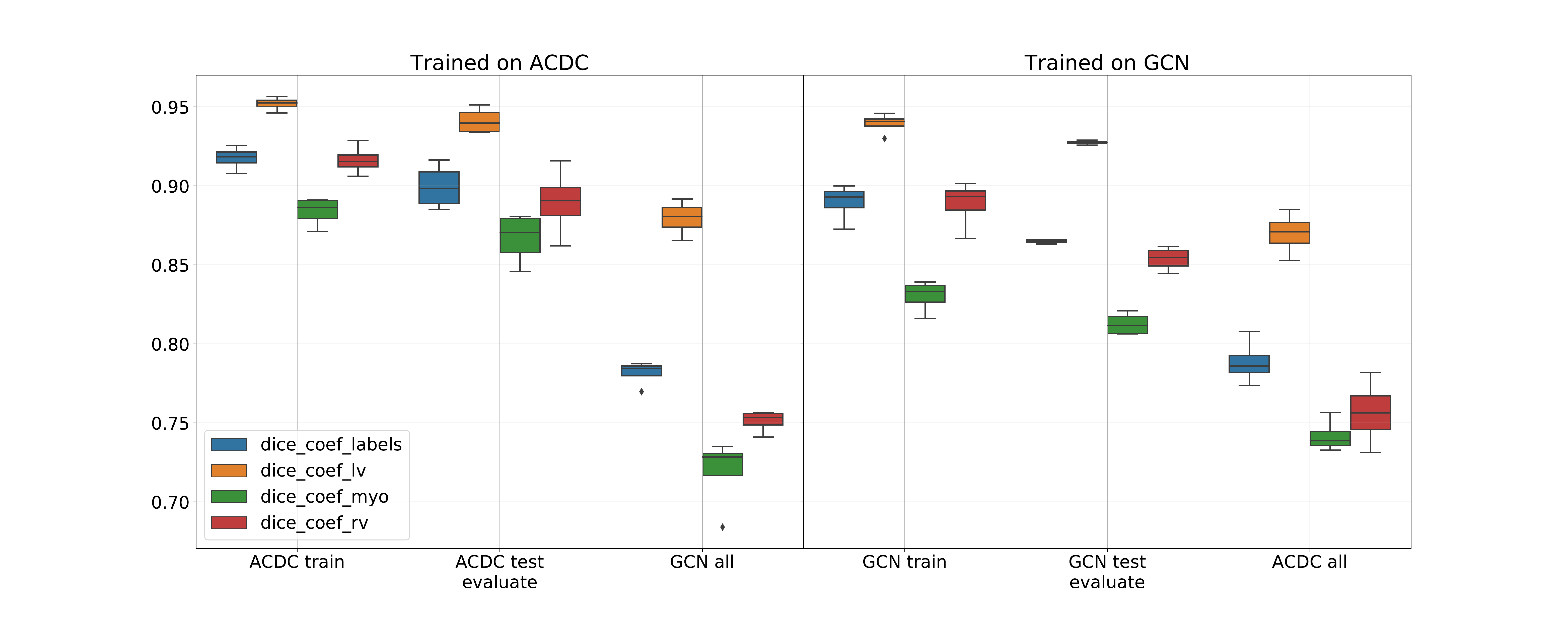}
  \caption[]{\label{fig:crossvalidation_box}Baseline dice scores of the cross-validated model performances trained on the split ACDC and GCN datasets. Each model is evaluated on the fold corresponding train/test split and the held out other dataset. The LV achieved the best dice scores across both datasets. The myocardium (MYO) seems to be the hardest label to learn, the RV dice is slightly better and achieved a remarkable small distribution within the GCN dataset for the ACDC models. The overall variation between the splits is very small which indicates that the splittings were good.}
\end{figure}

The following section defines the generalisation gap in both directions. Figure \ref{fig:crossvalidation_box} shows boxplots for the cross validated model performances within the four folds for both datasets. 
Table \ref{table:dice_scores} provides the mean dice scores per trainings- and evaluation-dataset and table \ref{table:dice_scores_gap} lists the absolute gap between the averaged training and evaluation scores.

To answer the question whether it is possible to train a U-net on a publicly available dataset and apply it afterwards on unseen pathologies, the following sections will mainly describe the model performances/ gaps.
The baseline models which were trained on the ACDC dataset achieved a mean $DSC_{labels}$ score of 0.917$\pm{0.006}$ /0.899$\pm{0.012}$ (train/test) and a mean train/test dice gap for all labels of 0.017 (0.026/0.010/0.016 for RV/LV/MYO). The LV (0.951$\pm{0.003}$/0.941$\pm{0.007}$ - train/test) seems to be the easiest label to learn. The trained models achieved a mean validation $DSC_{class}$ for the unseen test split of 0.941$\pm{0.007}$ for the LV, which is 0.027  below the ED score of the MICCAI winning model from Isensee et al. \cite{Isensee2018} and 0.010 $DSC_{class}$ better than the ES score of that winning model. The MYO seems to be the most difficult label to learn, related to the absolute test score of the ACDC models with a mean dice score of 0.866$\pm{0.014}$.

By validating these models on the unseen pathologies from the GCN dataset all $DSC_{class}$ scores dropped in a range from 0.072$\pm{0.001}$ to 0.165$\pm{0.001}$ compared to the training score. The dice score of the LV has the smallest drop (0.072$\pm{0.001}$) which confirms the expectations that the LV of TOF patients looks more like a normal LV. The dice score for the RV has the biggest drop with 0.165$\pm{0.001}$, which could be correlated with the clinical picture of TOF patients which usually have an deformed and increased end-systolic and end-diastolic volume of the right ventricle.

The models which were trained on the GCN dataset achieved lower train ($DSC_{labels}$ 0.889$\pm{0.010}$) and test dice scores ($DSC_{labels}$ 0.865$\pm{0.001}$) than the models trained on the ACDC dataset (0.917$\pm{0.006}$ / 0.899$\pm{0.012}$ - train/validation). On the other hand they were able to generalise better to the unseen dataset which results in lower $DSC_{labels}$ gaps for all labels (cf. table \ref{table:dice_scores_gap}). The reason for the smaller generalisation gap of the GCN-models could be the greater amount of CMR images and the greater clinical diversity of the GCN dataset, which forces the model to find other features.

\begin{table}[htb]
\centering
\caption{Baseline results regarding model generalisation. Dice scores averaged over the cross-validated (k=4) folds per training dataset. Evaluated on the corresponding train/test folds and the held out other dataset. The models achieved the best dice for the LV on all evaluation sets (marked in bold). The MYO had the lowest dice (0.866 $DSC_{class}$ - 0.812 $DSC_{class}$) for the test folds and the unseen dataset (0.719 $DSC_{class}$ - 0.741 $DSC_{class}$) across all experiments. The ACDC models performed overall better on their test sets than the GCN models. The best performing label is bold.}
\resizebox{.9\textwidth}{!}{%
\begin{tabular}{ccc|cccccc}
\\
Training & Evaluation & Evaluation  & Dice   & Dice   & Dice   & Dice   \\
dataset &  dataset &  modality &  Labels  &  RV  &  LV  &  MYO  \\
\hline
 & ACDC  & train & 0.917 $\pm{0.006}$ & 0.916 $\pm{0.008}$   & \textbf{0.951} $\pm{0.003}$  & 0.883 $\pm{0.008}$    \\
ACDC & ACDC  & test  & 0.899 $\pm{0.012}$ & 0.889 $\pm{0.019}$   & \textbf{0.941} $\pm{0.007}$  & 0.866 $\pm{0.014}$   \\
 & GCN   & all  & 0.781 $\pm{0.006}$ & 0.751 $\pm{0.006}$   & \textbf{0.879} $\pm{0.009}$   & 0.719 $\pm{0.020}$   \\
\hline
 & GCN   & train  & 0.889 $\pm{0.010}$ & 0.888 $\pm{0.009}$   & \textbf{0.939} $\pm{0.005}$  & 0.830 $\pm{0.008}$   \\
GCN & GCN   & test   & 0.865 $\pm{0.001}$ & 0.853 $\pm{0.006}$   & \textbf{0.927} $\pm{0.001}$  & 0.812 $\pm{0.006}$ \\
  & ACDC  & all  & 0.788 $\pm{0.012}$ & 0.756 $\pm{0.031}$   & \textbf{0.869} $\pm{0.011}$   & 0.741 $\pm{0.009}$
\end{tabular}%
}
\label{table:dice_scores}
\end{table}

\begin{table}[htb]
\centering
\caption{Model generalisation gap between the train/evaluation splits within a dataset and the train/unseen gap to the corresponding other dataset. The $DSC_{labels}$ gap between the training and test splits within the ACDC dataset were quite low (0.01 for the LV up to 0.026 for the RV). Even though the models generalised well within one dataset the overall gap to the unseen GCN dataset is bigger by a factor of seven  (0.017 - 0.136 $DSC_{labels}$). The ACDC models had a greater generalisation gap (0.3 $DSC_{labels}$) to the unseen dataset than the GCN models.}
\resizebox{.6\textwidth}{!}{%
\begin{tabular}{ccc|cccccc}
\\
Training & Evaluation & Evaluation  & Dice   & Dice   & Dice   & Dice   \\
dataset &  dataset &  modality &  Labels  &  RV  &  LV  &  MYO  \\
\hline
ACDC & ACDC  & test  & 0.017 & 0.026   & 0.010   & 0.016   \\
 & GCN   & all  & 0.136 & 0.165   & 0.072   & 0.164   \\
\hline
GCN & GCN   & test   & 0.024 & 0.017   & 0.011  & 0.034 \\
  & ACDC  & all  & 0.101 & 0.088   & 0.069   & 0.132
\end{tabular}%
}
\label{table:dice_scores_gap}
\end{table}

\subsection{Finetuning}
It turned out that all three methods closed the generalisation gap with nearly equal results for each of the labels and each of the added amount of GCN patients. The generalisation increase of the second method was slightly more stable and is shown in figure \ref{fig:close_gap_finetune}. The other improvement plots are attached in the appendix (cf. figure \ref{fig:close_gap_trainnew} and figure \ref{fig:close_gap_finetunegcn})).

\begin{figure}[ht]
\centering
  \includegraphics[width=.9\textwidth]{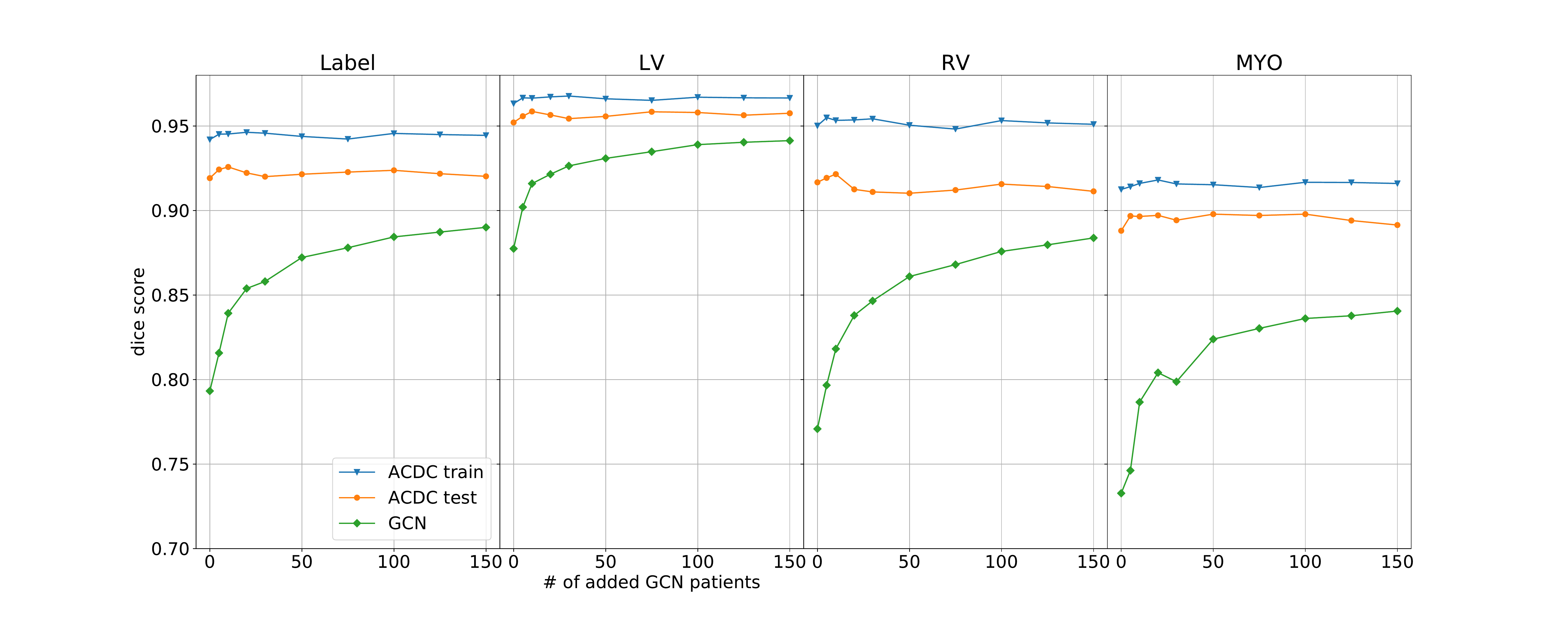}
  \caption[]{\label{fig:close_gap_finetune}Finetuning of the ACDC model with an increasing number of GCN patients. There is one plot for each label each with one line per dataset. The blue (upper) line describes the dice score on the ACDC split + n GCN patients, the orange (middle) line describes the dice score on the held out ACDC test split and the green line (lower) describes the dice score on the unseen GCN patients. The latter is exclusive the GCN patients which were added for the finetuning.}
\end{figure}

\begin{figure}[ht]
\centering
  \includegraphics[width=.8\textwidth]{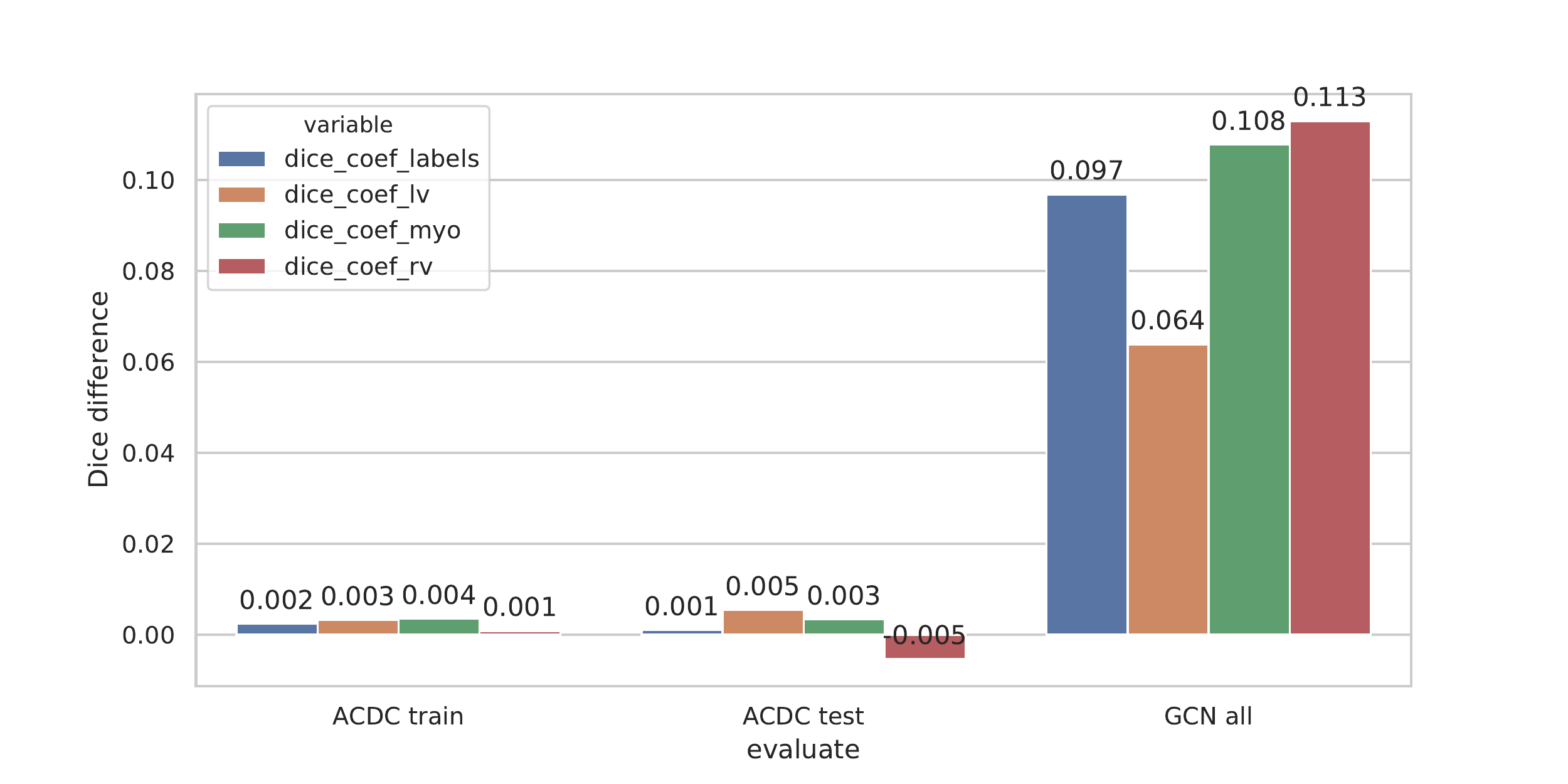}
  \caption[]{\label{fig:close_gap_increase}Dice score difference per label and modality by finetuning on additional 150 GCN patients. The model performance on the ACDC dataset did not drop due to the finetuning process, the performance on the GCN dataset increased by 0.10 $DSC_{labels}$.}
\end{figure}

Figure \ref{fig:close_gap_increase} illustrates the dice score increase between the baseline ACDC model and the best finetuned ACDC model (trained on ACDC + 150 GCN patients). All labels and each evaluation modality benefit from finetuning. The performance of the model on the GCN dataset increased by 0.10 $DSC_{labels}$, the ACDC train and test dice stayed stable. The RV and the MYO got the biggest boost (0.11 $DSC_{class}$), the left ventricle got the smallest boost (0.06 $DSC_{class}$) from finetuning the models.

\section{\hspace{14pt}Discussion}
\label{sec:disscussion}
The segmentation networks were able to segment the LV in the unseen dataset with a small reduction in the dice score (left ventricle dice score gap of 0.072 between the train dataset and the unseen pathology), but they generalised much worse to the TOF patients (right ventricle/ left ventricle myocardium dice score gap of 0.165/0.164).

This  indicates that the U-Net may still overfit to one dataset even if high dropout rates and massive data augmentation is applied. In this case the U-net performed very well within the 4-fold splits of the ACDC dataset and achieved a train/test $DSC_{labels}$ of 0.917$\pm{0.006}$/0.899$\pm{0.012}$  with a train/test gap of 0.017. The $DSC_{labels}$ dropped by 0.136 when the models were applied to patients with an unseen pathology, in this case TOF patients. The crosscheck (trained on GCN data, evaluated on ACDC data) showed a smaller but still mentionable gap between the two datasets (test/unseen $DSC_{labels}$ - 0.865$\pm{0.001}$/0.788$\pm{0.012}$) which supports the assumption of a pathology driven gap. Still it would be very interesting to see the performance of these models when they are applied to a healthy group of children. 

The results showed that this gap could be decreased by finetuning the ACDC model with the ACDC + TOF patients, which worked slightly better than continue the training of the baseline models only with the GCN data or training a new model with the ACDC + GCN data included from scratch. There are many other finetuning methods available, which might overcome the generalisation gap faster or with less necessary finetuning examples. One could try and finetune the models by fixing the weights of the decoding layers and finetune only the last encoding layers.

During the parameter optimisation it turned out that each loss function was able to train the models, but the learning process could be increased in convergence speed and accuracy a lot by choosing the right loss function for the given problem. All tested loss functions worked better if they are only applied to the foreground classes.

There is still the problem of rare pathologies with small available datasets, and the results indicate that it is necessary to finetune on them to make reasonable predictions.
Maybe the idea of federated and distributed learning as applied by Ken et al.\cite{Ken_distributed_learning_2018} could overcome the problem of rare datasets and pathologies by sharing the network weights across distributed institutions. 

\section{\hspace{14pt}Conclusion}
\label{sec:conclusion}

The experiments gave five mentionable insights. First, U-Nets are able to generalise very well to unseen data of the same pathology (with a gap of 1-2\% $DSC_{labels}$) if the training data is pre-processed (cf. section  \ref{list:preprocessing_steps}) and augmented in the right manner.
Second, U-Nets might overfit to the trained pathologies and generalise bad to unseen pathologies with deformed structures ($DSC_{labels}$ dropped by 0.136).
Third, the U-Net performance of the finetuned models was stable in segmenting the ACDC data, even when more than double of the trainings data (75 ACDC patients \& 150 GCN patients) consisted of patients with Tetralogy.
Fourth, the model performance on the TOF patients increased even after more than 125 TOF patients are added to the training data. One might expect that this converges earlier. The biggest performance gain could be generated by adding at least 50 patients.
Fifth, the generalisation gap towards the TOF pathology indicates that there are different features within the MRI images of TOF patients compared to the five pathologies contained in the ACDC data. This, together with the observation that the model performance did not stop to increase after a certain amount of patients, could be an indication that the TOF patients themselves have different features which could be used to define subgroups within the TOF patients.

%The current results indicate that there are still a lot of open questions related to the learning progress and feature selection of the U-Net which results in better or worse generalisation performance. 
%So far it is not clear how the model layers evolve while they finetune on a new pathology. It would be very interesting to define the layers with the biggest change due to a finetuning process. 

%Finally this work points that current deep learning models still need to be trained on all kind of physical deformed pathologies to make sure that they are still performing well. And that they can be used for surgical planning.
%Further research needs to be done to improve the network performance/generalisation and to create a time saving benefit for the daily clinical routine.

%flo Feedback 
%might expect
%might would overcome diss
%During the parameter

\section{\hspace{14pt}Acknowledgements}

The Titan Xp GPU card used for this research was donated by the NVIDIA Corporation.
This work was supported by the Competence Network for Congenital Heart Defects, which has received funding from the Federal Ministry of Education and Research, grant number 01GI0601 (until 2014), and the DZHK (German Centre for Cardiovascular Research; as of 2015).

% References
\bibliography{literatur.bib} 
\bibliographystyle{spiebib}
\clearpage
\appendix
%\section*{\hspace{14pt}Appendix}
\label{sec:appendix}

\begin{figure}
\centering
  \includegraphics[width=1.\textwidth]{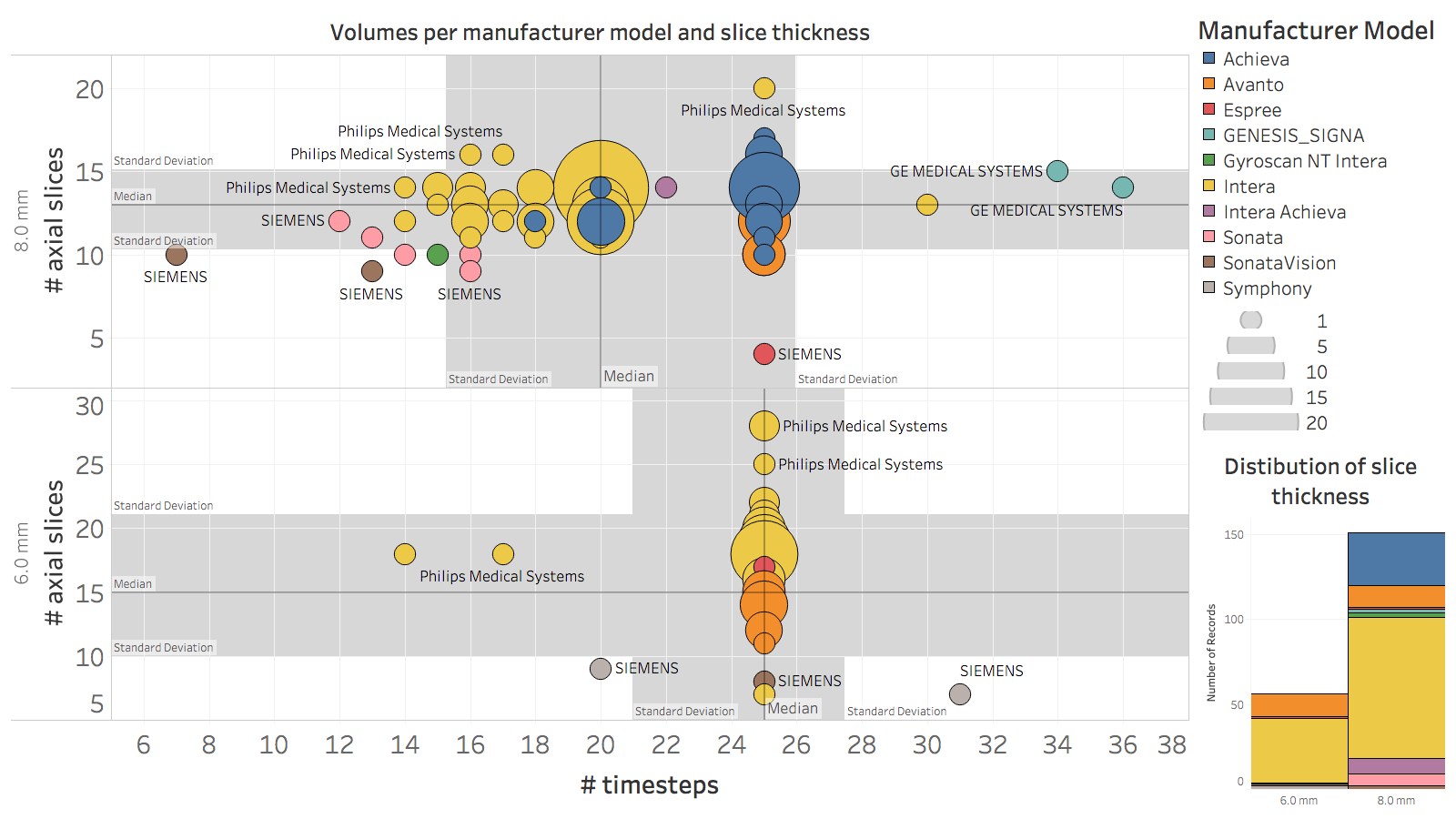}
  \caption{\label{fig:volumes_per_scanner}Distribution of the time- and axial number of slices per patient with labelled ground truth, grouped by manufacturer model. The x-axis represents the number of available time steps per patient. The y-axis shows the number of axial slices per patient and is split by the slice thickness of the CMR images (6mm, 8mm 10\&7mm). There was only one patient with a slice thickness of 7 mm and 10 mm. To improve the display of the others both are hidden on this plot. The median occurrence and the standard deviation of the number of z-slices per patient is shown by the grey stripes. The circle size reflects the number of patients with this spatial resolution.}
\end{figure}

\begin{figure}
\centering
  \includegraphics[width=1.\textwidth]{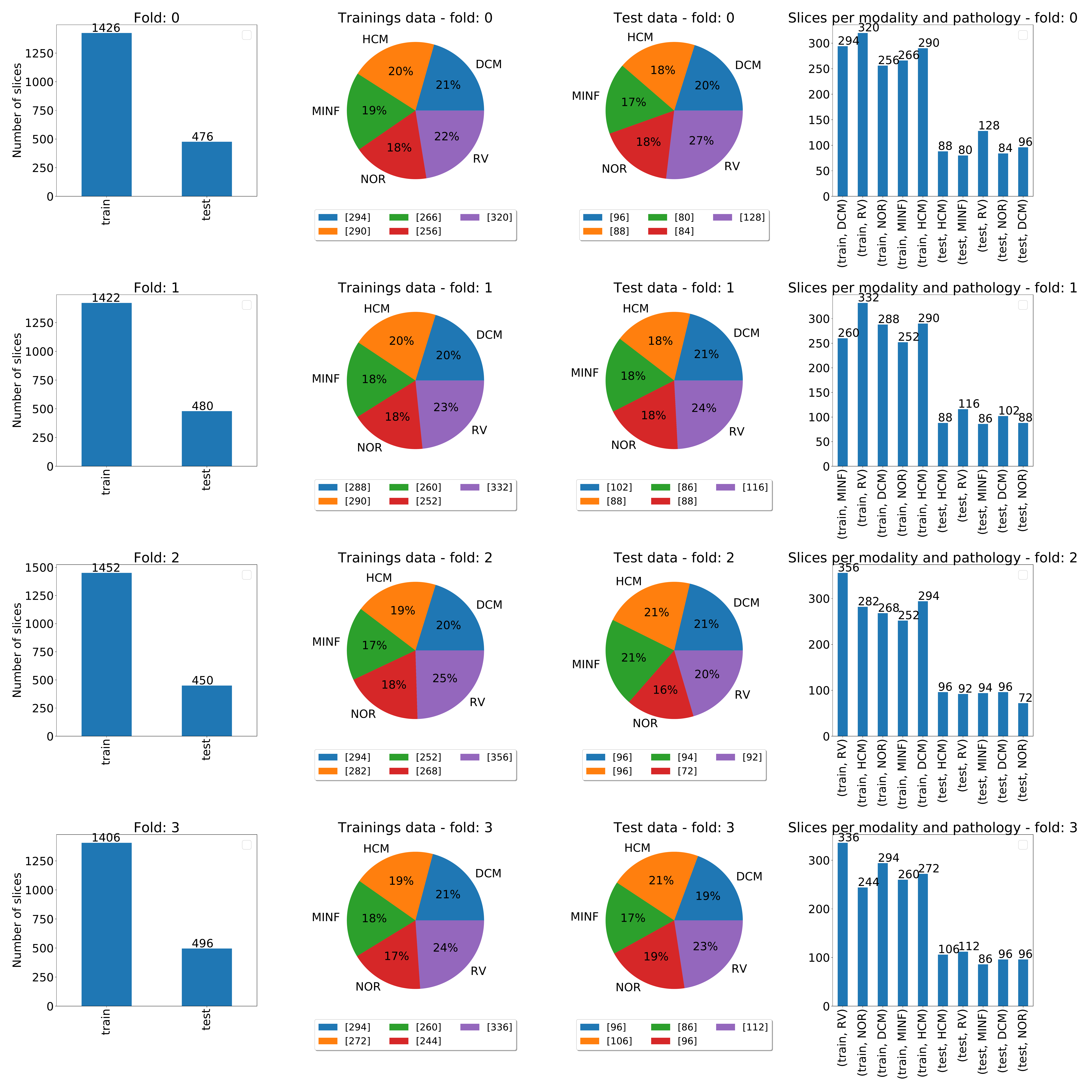}
  \caption{Overview of the split ACDC data. Each row represents one fold which leads to one trained model. The percentage differences in the pathology distribution between each fold comes from different number of slices on the z-axis. The ACDC dataset covers adults with normal cardiac anatomy and function (NOR) and the following four cardiac pathologies: systolic heart failure with infarction (MINF), dilated cardiomyopathy (DCM), hypertrophic cardiomyopathy (HCM) and abnormal right ventricular volume (ARV).}
  \label{fig:ACDC_data_splitting}
\end{figure}

\begin{figure}
\centering
  \includegraphics[width=1.\textwidth]{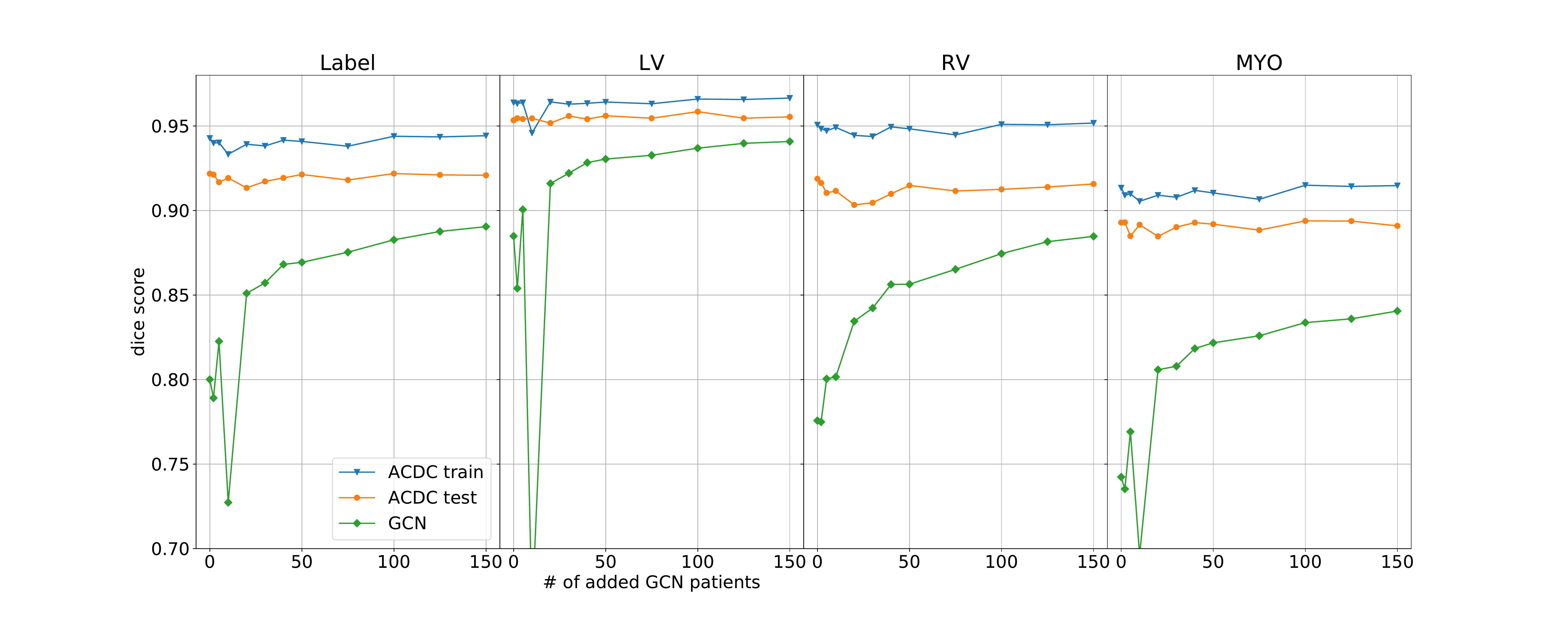}
  \caption{Finetuning of the ACDC model by training a new model with the ACDC train split and an increasing number of GCN patients. There is one plot for each label each with one line per dataset. The blue (upper) line describes the dice score on the ACDC split + n GCN patients, the orange (middle) line describes the dice score on the held out ACDC test split and the green line (lower) describes the dice score on the unseen GCN patients. The latter is exclusive the GCN patients which were added for the finetuning. The fourth model from left shows very bad results compared to the other finetuning models, which is caused by very bad parameter initialising. This experiment was not repeated to demonstrate one problem that might could occur, if only one model is trained and the model starts with unfortunately bad parameters.}
  \label{fig:close_gap_trainnew}
\end{figure}

\begin{figure}
\centering
  \includegraphics[width=1.\textwidth]{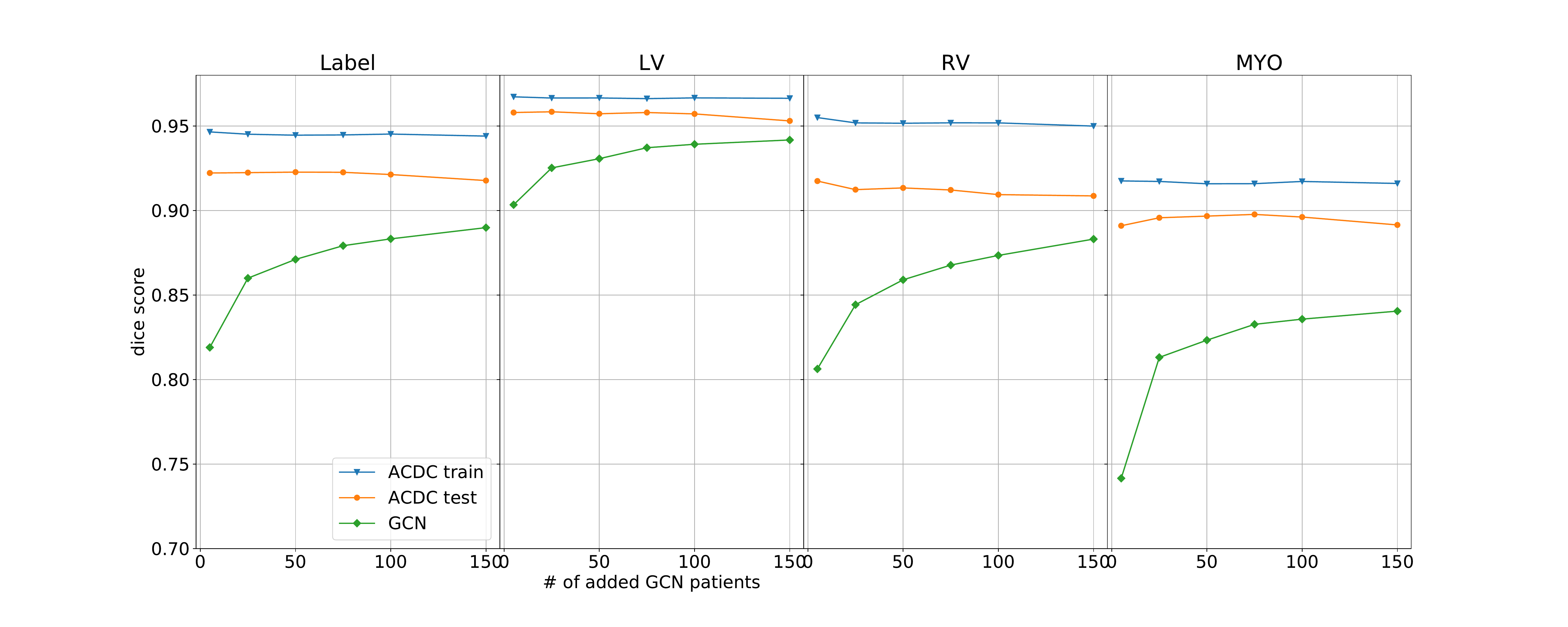}
  \caption{Finetuning of the ACDC model by continuing from the baseline with an increasing number of GCN patients. There is one plot for each label each with one line per dataset. The blue (upper) line describes the dice score on the ACDC split + n GCN patients, the orange (middle) line describes the dice score on the held out ACDC test split and the green line (lower) describes the dice score on the unseen GCN patients. The latter is exclusive the GCN patients which where added for the finetuning.}
  \label{fig:close_gap_finetunegcn}
\end{figure}

\clearpage

\end{document}